\newcommand{\CLASSINPUTtoptextmargin}{0.75in}%
\newcommand{\CLASSINPUTbottomtextmargin}{1in}%
\newcommand{\CLASSINPUTinnersidemargin}{0.63in}%
\newcommand{\CLASSINPUToutersidemargin}{0.63in}%
\documentclass[conference,10pt,letterpaper]{IEEEtran}%
\usepackage{amsmath}%
\usepackage{graphicx}%
\usepackage{multirow}%
\usepackage[none]{hyphenat}%
\usepackage{float}%
\usepackage{subfig}%
\usepackage{dblfloatfix}%
\usepackage{t1enc}%
\usepackage{times}
\usepackage{cite}
\usepackage[flushleft]{threeparttable}
\usepackage{url}
\usepackage{amsmath,mathtools,color}
\usepackage{xcolor,soul,framed}
\usepackage{makecell}
\usepackage{lipsum}
\usepackage{multicol}
\usepackage{xcolor,cite,etoolbox}
\usepackage{footnote}
\usepackage{multirow}
\usepackage{orcidlink}
\makeatletter

\def\@IMSauthorblockNAMEstyle{\normalfont\IMSauthorsize}
\def\@IMSauthorblockAFFILstyle{\normalfont\IMSaffilsize}
\def\@IMSauthorblockEMAILstyle{\normalfont\IMSaffilsize}
\def\IMSauthorblockNAME#1{%
\relax\@IMSauthorblockNAMEstyle%
#1%
}%
\def\IMSauthorblockAFFIL#1{%
\relax\@IMSauthorblockAFFILstyle%
\vskip\@IEEEauthorblockAtopspace
#1%
}%
\def\IMSauthorblockEMAIL#1{%
\relax\@IMSauthorblockEMAILstyle%
\vskip\@IEEEauthorblockAtopspace
#1%
}%
\newcommand{\IMSauthor}[1]{%
\ifIsBlindReviewVersion%
\author{\phantom{\parbox{\textwidth}{\center\relax#1}}}%
\else%
\author{\parbox{\textwidth}{\center\relax#1}}%
\fi%
}%
\newif\ifIsBlindReviewVersion

\def\@maketitle{\newpage
\bgroup\par\addvspace{0.5\baselineskip}\centering%
\ifCLASSOPTIONtechnote
   {\bfseries\large\@IEEEcompsoconly{\sffamily}\@title\par}\vskip 1.3em{\lineskip .5em\@IEEEcompsoconly{\sffamily}\@author
   \@IEEEspecialpapernotice\par{\@IEEEcompsoconly{\vskip 1.5em\relax
   \@IEEEtitleabstractindextextbox{\@IEEEtitleabstractindextext}\par
   \hfill\@IEEEcompsocdiamondline\hfill\hbox{}\par}}}\relax
\else
   \vskip0.2em{\IMStitlesize\ifCLASSOPTIONtransmag\bfseries\LARGE\fi\@IEEEcompsoconly{\sffamily}\@IEEEcompsocconfonly{\normalfont\normalsize\vskip 2\@IEEEnormalsizeunitybaselineskip
   \bfseries\Large}\@title\par}\vskip1.0em\par
   \ifCLASSOPTIONconference%
      {\@IEEEspecialpapernotice\mbox{}\vskip\@IEEEauthorblockconfadjspace%
       \mbox{}\hfill\begin{@IEEEauthorhalign}\@author\end{@IEEEauthorhalign}\hfill\mbox{}\par}\relax
   \else
      \ifCLASSOPTIONpeerreviewca
         {\@IEEEcompsoconly{\sffamily}\@IEEEspecialpapernotice\mbox{}\vskip\@IEEEauthorblockconfadjspace%
          \mbox{}\hfill\begin{@IEEEauthorhalign}\@author\end{@IEEEauthorhalign}\hfill\mbox{}\par
          {\@IEEEcompsoconly{\vskip 1.5em\relax
           \@IEEEtitleabstractindextextbox{\@IEEEtitleabstractindextext}\par\hfill
           \@IEEEcompsocdiamondline\hfill\hbox{}\par}}}\relax
      \else
         \ifCLASSOPTIONtransmag
           {\@IEEEspecialpapernotice\mbox{}\vskip\@IEEEauthorblockconfadjspace%
            \mbox{}\hfill\begin{@IEEEauthorhalign}\@author\end{@IEEEauthorhalign}\hfill\mbox{}\par
           {\vspace{0.5\baselineskip}\relax\@IEEEtitleabstractindextextbox{\@IEEEtitleabstractindextext}\vspace{-1\baselineskip}\par}}\relax
         \else
           {\lineskip.5em\@IEEEcompsoconly{\sffamily}\sublargesize\@author\@IEEEspecialpapernotice\par
           {\@IEEEcompsoconly{\vskip 1.5em\relax
            \@IEEEtitleabstractindextextbox{\@IEEEtitleabstractindextext}\par\hfill
            \@IEEEcompsocdiamondline\hfill\hbox{}\par}}}\relax
         \fi
      \fi
   \fi
\fi\par\addvspace{0.0\baselineskip}\egroup}

\def\IMStitlesize{\@setfontsize{\IMStitlesize}{18}{21pt}}
\def\IMSauthorsize{\@setfontsize{\IMSauthorsize}{12}{13pt}}
\def\IMSaffilsize{\@setfontsize{\IMSaffilsize}{12}{13pt}}
\def\IMScaptionsize{\@setfontsize{\IMScaptionsize}{8}{9pt}}
\def\IMSbibsize{\@setfontsize{\IMSbibsize}{8}{9pt}}

\def\@IEEEauthorblockNstyle{\IMSauthorsize\@IEEEcompsocnotconfonly{\sffamily}\@IEEEcompsocconfonly{\large}}
\def\@IEEEauthorblockAstyle{\IMSaffilsize\@IEEEcompsocnotconfonly{\sffamily}\@IEEEcompsocconfonly{\itshape}\@IEEEcompsocconfonly{\large}}
\def\@IEEEauthordefaulttextstyle{\IMSauthorsize\@IEEEcompsocnotconfonly{\sffamily}\sublargesize}

\def\thebibliography#1{\section*{\refname}%
    \addcontentsline{toc}{section}{\refname}%
    \IMSbibsize\@IEEEcompsocconfonly{\small}\vskip 0.3\baselineskip plus 0.1\baselineskip minus 0.1\baselineskip
    \list{\@biblabel{\@arabic\c@enumiv}}%
    {\settowidth\labelwidth{\@biblabel{#1}}%
    \leftmargin\labelwidth
    \advance\leftmargin\labelsep\relax
    \itemsep \IEEEbibitemsep\relax
    \usecounter{enumiv}%
    \let\p@enumiv\@empty
    \renewcommand\theenumiv{\@arabic\c@enumiv}}%
    \let\@IEEElatexbibitem\bibitem%
    \def\bibitem{\@IEEEbibitemprefix\@IEEElatexbibitem}%
\def\newblock{\hskip .11em plus .33em minus .07em}%
\ifCLASSOPTIONtechnote\sloppy\clubpenalty4000\widowpenalty4000\interlinepenalty100%
\else\sloppy\clubpenalty4000\widowpenalty4000\interlinepenalty500\fi%
    \sfcode`\.=1000\relax}

\long\def\@makecaption#1#2{%
\ifx\@captype\@IEEEtablestring%
\par\@IEEEtabletopskipstrut
\else
\@IEEEfigurecaptionsepspace
\fi
\setbox\@tempboxa\hbox{\normalfont\IMScaptionsize {#1.}\nobreakspace\nobreakspace #2}%
\ifdim \wd\@tempboxa >\hsize%
\setbox\@tempboxa\hbox{\normalfont\IMScaptionsize {#1.}\nobreakspace\nobreakspace}%
\parbox[t]{\hsize}{\normalfont\IMScaptionsize\noindent\unhbox\@tempboxa#2}%
\else
\ifCLASSOPTIONconference \hbox to\hsize{\normalfont\IMScaptionsize\hfil\box\@tempboxa\hfil}%
\else \hbox to\hsize{\normalfont\IMScaptionsize\box\@tempboxa\hfil}%
\fi\fi
\ifx\@captype\@IEEEtablestring%
\@IEEEtablecaptionsepspace
\else
\fi}

\newlength\tablecaptiontotableskip
\newlength\figuretocaptionskip
\def\@IEEEfigurecaptionsepspace{\vskip\figuretocaptionskip\relax}%
\def\@IEEEtablecaptionsepspace{\vskip\tablecaptiontotableskip\relax}%

\def\abstract{\normalfont%
\@IEEEabskeysecsize\bfseries\textit{\abstractname}\,\bfseries\textit{---}\,%
\@IEEEgobbleleadPARNLSP}%

\def\IEEEkeywords{\normalfont%
\@IEEEabskeysecsize\bfseries\textit{\IEEEkeywordsname}\,\bfseries\textit{---}\,%
\@IEEEgobbleleadPARNLSP}%
\def\endIEEEkeywords{\relax\vspace{0.67ex}%
\par\if@twocolumn\else\endquotation\fi%
\normalsize\normalfont}%

\DeclareRobustCommand*{\IMSauthorrefmark}[1]{\raisebox{0pt}[0pt][0pt]{\textsuperscript{\footnotesize{#1}}}}%
\def\@IEEEauthorblockNtopspace{0ex}
\def\@IEEEauthorblockAtopspace{1mm}
\def\tablename{Table}
\def\thetable{\arabic{table}}
\def\IEEEkeywordsname{Keywords}
\def\subsubsection{\@startsection{subsubsection}{3}{\z@}{1.5ex plus 1.5ex minus 0.5ex}%
{0.7ex plus .5ex minus 0ex}{\normalfont\normalsize\itshape}}%
\def\@seccntformat#1{\csname the#1dis\endcsname\relax}
\def\thesectiondis{\thesection.\hskip 0.5em}
\def\thesubsectiondis{{\hbox to\parindent{\Alph{subsection}.}}}
\def\thesubsubsectiondis{{\hbox to \parindent{\arabic{subsubsection})}}}
\def\theparagraphdis{{\hbox to \parindent{\alph{paragraph})}}}
\IEEEilabelindentA \parindent
\IEEEilabelindent \IEEEilabelindentA
\IEEEelabelindent \parindent
\IEEEdlabelindent \parindent
\IEEElabelindent \parindent

\newlength\@IMSparindent
\newcommand\IMSdisplayacksection[1]{%
\ifIsBlindReviewVersion%
\noindent\phantom{\parbox[t]{\columnwidth}{\normalbaselines\setlength{\parindent}{\@IMSparindent}{#1}\strut}}
\else%
\noindent\parbox[t]{\columnwidth}{\normalbaselines\setlength{\parindent}{\@IMSparindent}{#1}\strut}%
\fi%
}%

\makeatother

\usepackage{lipsum}
\usepackage{fancyhdr}
\fancypagestyle{sec}{\lhead{\tmpx}}

\fancypagestyle{arXiv}
{
   \fancyhf{}
   \chead{\textcolor{gray}{This article was presented at the IEEE MTT-S International Microwave Symposium (IMS 2024),\\ Washington, DC, USA, June 16–21, 2024.}}
   \cfoot{ \fontsize{=9pt}{10pt}\selectfont  \textcolor{gray}{© 2024 IEEE. Personal use of this material is permitted. Permission from IEEE must be obtained for all other uses, in any current or future media, including reprinting/republishing this material for advertising or promotional purposes, creating new collective works, for resale or redistribution to servers or lists, or reuse of any copyrighted component of this work in other works.}\\\fontsize{=10pt}{12pt}\selectfont\thepage}
   
}
\makeatletter
\patchcmd{\@maketitle}
  {\addvspace{0.5\baselineskip}\egroup}
  {\addvspace{-1\baselineskip}\egroup}
  {}
  {}
\makeatother
\begin{document}
\raggedbottom

\title{Transfer Learning Assisted Fast Design Migration Over Technology Nodes: A Study on Transformer Matching Network }

\IMSauthor{%
\IMSauthorblockNAME{
Chenhao Chu\orcidlink{0000-0002-2562-9599},
Yuhao Mao\orcidlink{0000-0002-9803-8669},
Hua Wang\orcidlink{0000-0003-4952-5505}
}
\\%
\IMSauthorblockAFFIL{
\IMSauthorrefmark{}ETH Z\"urich, Switzerland
}
\\%
\IMSauthorblockEMAIL{
\IMSauthorrefmark{}chenhao.chu@iis.ee.ethz.ch,
\IMSauthorrefmark{}yuhao.mao@inf.ethz.ch,
\IMSauthorrefmark{}huawang@ethz.ch
}
}

\maketitle

\begin{abstract}

In this study, we introduce an innovative methodology for the design of mm-Wave passive networks that leverages knowledge transfer from a pre-trained synthesis neural network (NN) model in one technology node and achieves swift and reliable design adaptation across different integrated circuit (IC) technologies, operating frequencies, and metal options. We prove this concept through simulation-based demonstrations focusing on the training and comparison of the coefficient of determination ($R^2$) of synthesis NNs for 1:1 on-chip transformers in GlobalFoundries(GF) 22nm FDX+ (target domain), with and without transfer learning from a model trained in GF 45nm SOI (source domain). 
In the experiments, we explore varying target data densities of 0.5\%, 1\%, 5\%, and 100\% with a complete dataset of 0.33 million in GF 22FDX+, and for comparative analysis, apply source data densities of 25\%, 50\%, 75\%, and 100\% with a complete dataset of 2.5 million in GF 45SOI. With the source data only at 30GHz, the experiments span target data from two metal options in GF 22FDX+ at frequencies of 30 and 39 GHz. The results prove that the transfer learning with the source domain knowledge \mbox{(GF 45SOI)} can both accelerate the training process in the target domain (GF 22FDX+) and improve the $R^2$ values compared to models without knowledge transfer. Furthermore, it is observed that a model trained with just 5\% of target data and augmented by transfer learning achieves $R^2$ values superior to a model trained with 20\% of the data without transfer, validating the advantage seen from 1\% to 5\% data density. This demonstrates a notable reduction of 4X in the necessary dataset size highlighting the efficacy of utilizing transfer learning to mm-Wave passive network design. The PyTorch learning and testing code is publicly available at \textcolor{blue}
{\url{https://github.com/ETH-IDEAS/RFIC-TL}}.
\end{abstract}

%
%

\section{Introduction}\label{intro}
\begin{figure}[tp]
\centering
\includegraphics[width=0.8\linewidth]{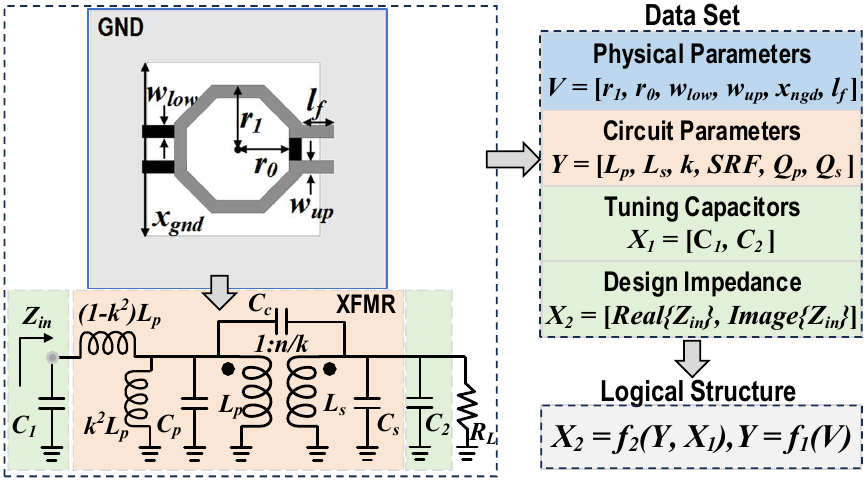}
\caption{Basic information about the transformer-based impedance matching network: EM structure, schematic structure, and parameter descriptions.}
\label{xfmr_details}
\vspace{-1.2em}
\end{figure}

\thispagestyle{arXiv}
The evolution of wireless communication brings challenges to the circuit and system design, with passive networks playing a crucial role in achieving optimal performance. Traditional methods involve iterative tuning and optimization through electromagnetic (EM) simulators. As illustrated in Fig. \ref{xfmr_details}, designing an on-chip transformer with desired design impedance \textit{$X_2$} and tuning capacitance \textit{$X_1$}, requires building a 3D EM structure using physical parameters \textit{$V$}, extracting \textit{$Y$} from EM simulations, and verifying \textit{$X_2$} via synthesis schematics. In particular, these approaches are often laborious and less effective when migrating designs across different technology nodes, which often involves changes in metal options besides changing foundry integrated circuit (IC) processes. Therefore, there is an unmet need for innovative approaches to radically accelerate design migrations across technology nodes to facilitate rapid time-to-market prototyping and productization.

Deep neural networks (DNNs) have been increasingly applied in the forward and inverse design of passive networks \cite{Jiang2021}. The DNN-assisted methods for synthesis of on-chip EM structures report the ability to predict physical parameters based on the given circuit parameters \cite{MunzerRFIC2020} and optimal input impedance \cite{er2021deep}. The NN-based automated matching circuit synthesizer shows advanced capability in providing proper circuit layout parameters at different frequencies \cite{LeeAMCS2022}. 

\begin{figure*}[tp]
\centering
\includegraphics[width=0.85\linewidth]{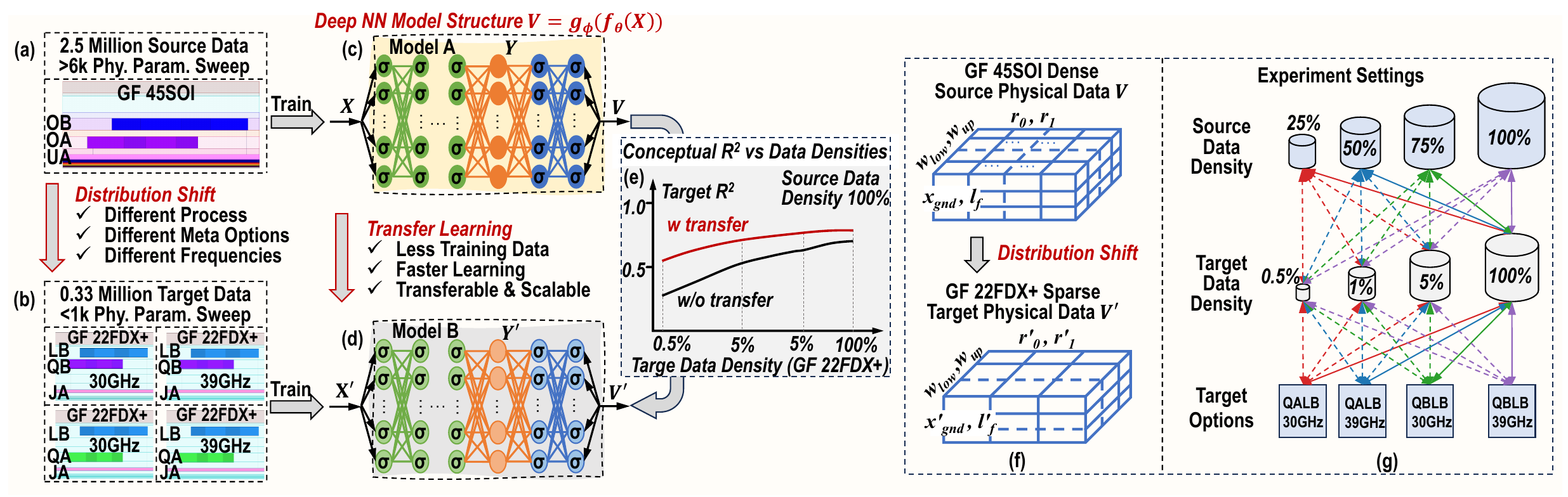}
\caption{Conceptual diagram of the proposed transfer learning approach across different IC technologies. (a) Source: 2.8 million data in GF 45SOI; (b) Target: 0.33 million data in GF 22FDX+; (c) Model structure for the proposed NN-based synthesizer trained from source data; (d) Knowledge-transferred model with target data; (e) Conceptual $R^2$ vs target data density with/without transfer; (f) Description of grid data distribution shift from source to target. (g) Experiment settings with different target and source data densities for two GF 22FDX+ metal options at 30 and 39 GHz.}
\label{transfer_learning_details}
\vspace{-1.5em}
\end{figure*}

However, the advancements in the IC "specs-to-layout" challenge still heavily depend on EM simulators for accurate training data, which inevitably leads to high computational costs. As on-chip transformers are governed by physical and EM principles, data generation often favors parameter sweeping over random searches, requiring insight into physical parameters. Furthermore, models pre-trained for one technology node in Fig. \ref{transfer_learning_details}(a), are not readily applicable to passive designs in others shown in Fig. \ref{transfer_learning_details}(b). Given these constraints, the efficiency of NN-based synthesizers may not necessarily outperform traditional methods when migrating designs across different technology nodes.

To overcome these challenges, we introduce an innovative approach that leverages knowledge transfer from a pre-trained synthesis NN model to achieve swift and reliable design adaptation across different IC technologies, operating frequencies, and metal options. Fig. \ref{transfer_learning_details}(c) and (d) demonstrate our approach, which relies on adapting models through transfer learning. We aim to reduce the need for extensive data collection for the target task. We aspire to achieve superior or comparable $R^2$ results from models using sparse target data through transfer learning, in contrast to models without transfer learning but with complete target data, as shown in Fig. \ref{transfer_learning_details}(e). We validate this methodology by evaluating the coefficient of determination ($R^2$) in synthesis NNs for 1:1 on-chip transformers at different data densities in \mbox{GlobalFoundries (GF) 22nm FDX+ (target)} shown in \mbox{Fig. \ref{transfer_learning_details}(b)}, with and without transfer learning from a pre-trained in \mbox{GF 45nm SOI} (source), as depicted in \mbox{Fig. \ref{transfer_learning_details}(a)}.


\section{Transfer Learning Approach} \label{sec:method}

As shown in \mbox{Fig. \ref{xfmr_details}}, the geometry structure of on-chip transformers are governed by physical and EM principles, and the data generation of physical parameters \textit{$V$} often favors parameter sweeping over random searches. As the mapping from transformer geometry to circuit parameters, \textit{$Y$} has the same data distribution with \textit{$V$}. The data of desired impedance \textit{$X_2$} is generated by calculating \textit{$Y$} with the tuning capacitance values \textit{$X_1$}, which are also swept in the defined range and step. These indicate the natural characteristic of the grid structure feature for data generation of the on-chip transformer. However, dealing with high-dimensional grids is challenging because the total number of potential combinations grows exponentially with each additional dimension. While it is possible to gather a lot of data within a specific grid for training machine learning models, applying these models to a new grid is difficult due to changes in the data distribution, which is illustrated in Fig. \ref{transfer_learning_details}(f). This shift in data distribution can be caused by factors like different technologies, metal options, and frequencies, which weaken the performance of the original model. Moreover, gathering an equivalent amount of data for each new grid is often too expensive, making it hard to develop models that are specifically tailored for these new grids.

\subsection{Source and Target Grid Data} 2.5 million source data was generated for GF 45SOI, with >6k physical sweeping in Ansys HFSS. For each target option in GF 22FDX+, as elaborated in Fig. \ref{transfer_learning_details}(g), 0.33 million data was generated with <1k physical sweeping. For training, we extract the mean and divide it by the corresponding standard deviation for each data grid. Further, we divide the full dataset into train, validation, and test sub-datasets with the ratio 6:2:2.

\subsection{Model Architecture} Fig. \ref{xfmr_details} demonstrates the logical structures, where the physical parameters \textit{$V$} decide the circuit parameters \textit{$Y$}, which along with inputs from tuning capacitors \textit{$X_{1}$}, determine the design impedance \textit{$X_{2}$}. Fig. \ref{transfer_learning_details}(b) shows the design of the synthesis network, tailored to directly predict the physical parameters \textit{$V$} from specified inputs \textit{$X_{1}$} and \textit{$X_{2}$}. The model structure begins with a circuit net predicting \textit{$Y$}, followed by a physical net that calculates \textit{$V$} based on the circuit net’s outputs. Throughout the paper, we use multi-layer perceptron (MLP) \cite{haykin1994neural} where all linear layers except the last one are followed by a batch norm layer to accelerate training and improve performance \cite{IoffeS15}. The circuit net is a 7-layer MLP and the physical net is a 3-layer MLP. All hidden dimensions are set to 512 and the activation function is ReLU \cite{agarap2018deep}.

\subsection{Loss Function} We combine the information from \textit{$Y$} and \textit{$V$} by training the model with the following loss function:
\begin{equation} \label{eq:loss}
    L_{\theta, \phi}(x, y, v) = \tau (y - f_\theta(x))^2 + (v - g_\phi \circ f_\theta(x))^2
\end{equation}
where $f_\theta$ is the circuit net, $g_\phi$ is the physical net and $\tau=0.5$ controls the weight for circuit information. To evaluate the performance of the model, we apply $R^2$ averaged over all output dimensions as the criterion, defined as
\begin{equation}
    R^2 = \frac{1}{K} \sum_{i=1}^K \left( 1 - \frac{\sum_{j=1}^N (v^j_i - g_\phi \circ f_\theta (x^j_i))^2}{\sum_{j=1}^N (v^j_i - \bar{v}_i)^2}\right)
\end{equation}
where $i$ represents the iterations over all output dimensions, $j$ iterates over the dataset and $\bar{v}_i$ is the mean of $v^j_i$ over $j$. This metric is invariant to the standard deviation and mean of each output dimension in the test data, assigning equal weight to every output dimension.

\subsection{Training Procedure} We use Adam \cite{KingmaB14Adam}, a state-of-the-art optimizer, to train the model using \eqref{eq:loss}. Learning rate is set to $5\times 10^{-4}$ and weight decay is set to $10^{-4}$. The model is trained for 300 epochs with a batch size equal to 4096, and the learning rate is multiplied by $0.2$ every 50 epochs after epoch 150. We train every model following the same procedure for fair comparison.

We aim to convey knowledge from the source grid, which has a relatively dense dataset, to the target grid where the dataset is much sparser. We start by training a model on the source grid data and then refine it with the target grid data. Given the distribution shift between the source and target, we keep the entire model adaptable, allowing for full adjustment to the target distribution. The effectiveness of this transfer learning is assessed by calculating the gain in terms of relative improvement (RI), which we define as:
\begin{equation}
    \text{RI} (\%) = \frac{R^2_{\text{transfer}} - R^2_{\text{non-transfer}}}{R^2_{\text{non-transfer}}} \cdot 100\% 
\end{equation}

\section{Simulation Experiments}

\begin{figure}[tp]  
\centering
\includegraphics[width=0.4\linewidth]{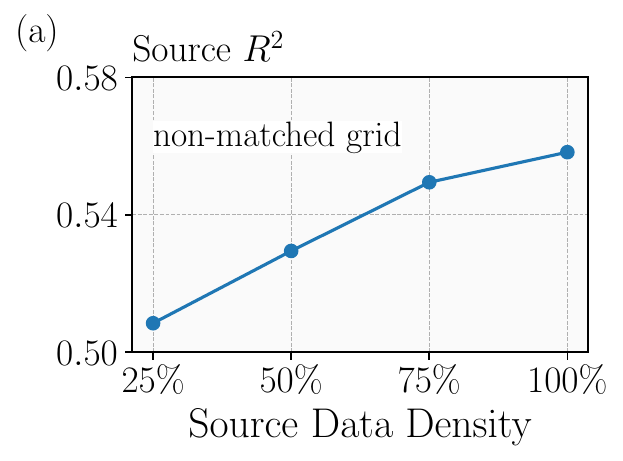}
\hfil
\includegraphics[width=0.4\linewidth]{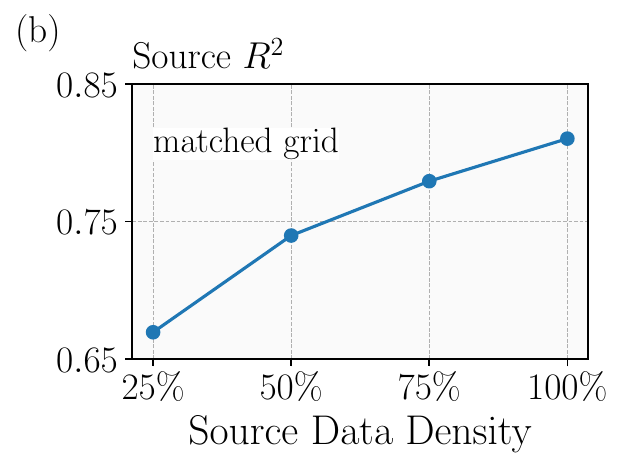}
\caption{Source $R^2$ versus Source Data Density: (a) $R^2$ on a larger, non-matched source grid; (b) $R^2$ on a smaller, matched grid, with the latter's higher $R^2$ due to its simplicity and fewer learning possibilities.}
\label{src}
\vspace{-1em}
\end{figure}

To validate the impact of transfer learning, we present simulation experimental results. As shown in Fig. \ref{transfer_learning_details}(g), we apply knowledge transfer from models trained on \mbox{GF 45SOI} OAOB 30GHz grid data to models designed for four distinct grids in GF 22FDX+: QALB 30GHz, QALB 39GHz, QBLB 30GHz, and QBLB 39GHz. To examine the impact of source data density, we train four different source models with 25\%, 50\%, 75\%, and 100\% data sampled uniformly at random from the complete source grid, respectively. The performances of these source models are depicted in Fig. \ref{src}(a). As these models are evaluated on the identical data distribution, it is observed that increased data density leads to enhanced model performance. We then employ the proposed transfer learning approach from the source models to different target grids.

\begin{figure}[tp]  
\centering
\includegraphics[width=0.45\linewidth]{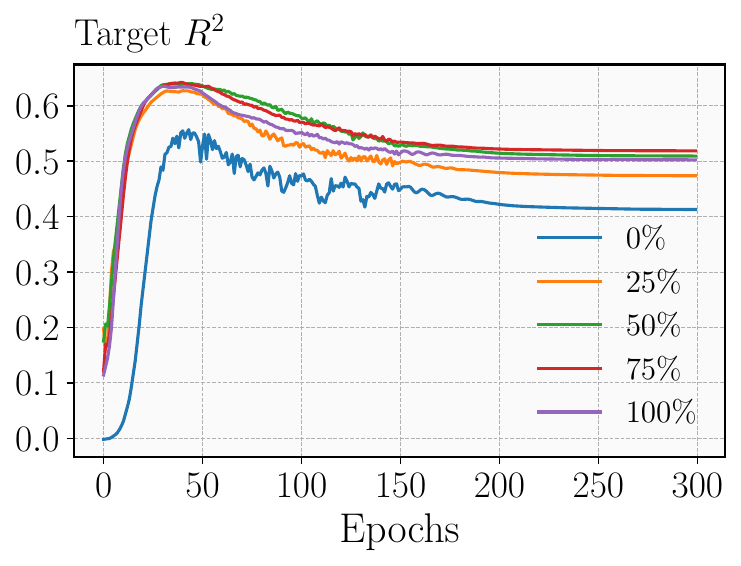}
\caption{Comparison of training dynamics: models trained with diverse source data densities on validation set, with 0\% source density indicating training exclusively on target data (without transfer).}
\label{dynamics}
\vspace{-1.5em}
\end{figure}

Fig. \ref{dynamics} shows training dynamics from four source models to GF 22FDX+ QALB 30GHz grid with 1\% data density, against models without knowledge transfer (blue). Transfer learning accelerates the training process and, more importantly, enhances performance and is more robust to overfitting. This proves that transfer learning can effectively improve the model on an unseen grid with highly sparse data.

\begin{figure}[tp]  
\centering
\includegraphics[width=0.8\linewidth]{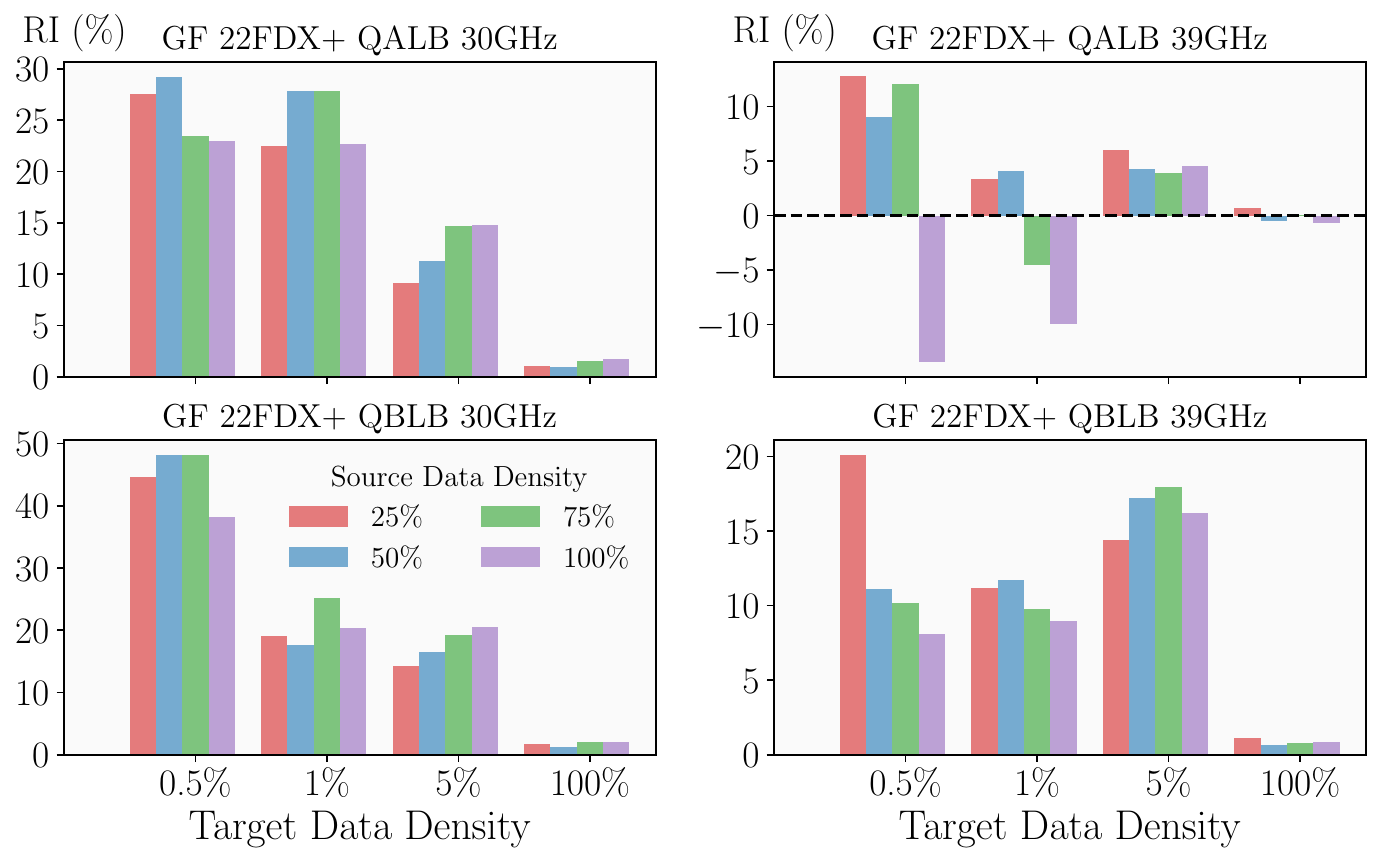}
\caption{Relative improvement (RI) versus source and target data densities for four target grid variations.}
\label{RI_full}
\vspace{-1.5em}
\end{figure}

Further, we discuss the effect of target data density and source data density. We use four different levels of target data density: 0.5\%, 1\%, 5\%, and 100\%. The first three represent different densities in highly sparse data, while the last one aims to examine whether transfer learning is still useful with complete data in the target grid. The RI of transfer learning for these settings is shown in Fig. \ref{RI_full}. 
First, we observe that in almost all settings, including those with dense target data, transfer learning exhibits better performance, as the RI is positive. Second, while the RI in the complete target data case is around 1\%, the effect becomes increased when the data is sparse, e.g., around 20\% for 5\% density and close to 50\% for 0.5\% density in the bottom left of Fig. \ref{RI_full}. This confirms our intuition that transfer learning assists more when less data is available in the target grid. Third, due to grid knowledge mismatch,
a source model trained with less data in the source grid may offer more assistance to the target model than one trained with more data. For example, the results of GF 22FDX+ QALB 30GHz show that the source model with 25\% train data provides more assistance than the source model with 100\% train data. This rule is general, since in most cases, appropriate source models outperform those trained with a complete dataset when applied to the target grid.

\begin{figure}[tp]  
\centering
\includegraphics[width=0.4\linewidth]{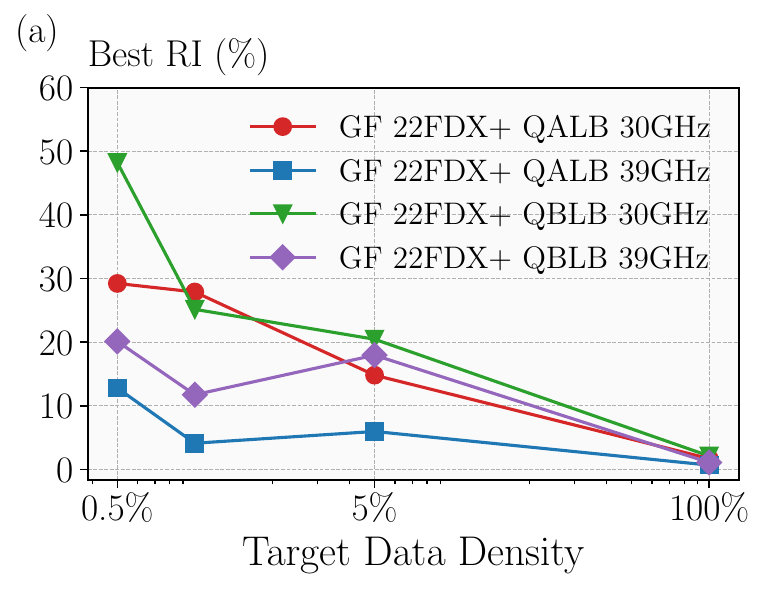}
\hfil
\includegraphics[width=0.4\linewidth]{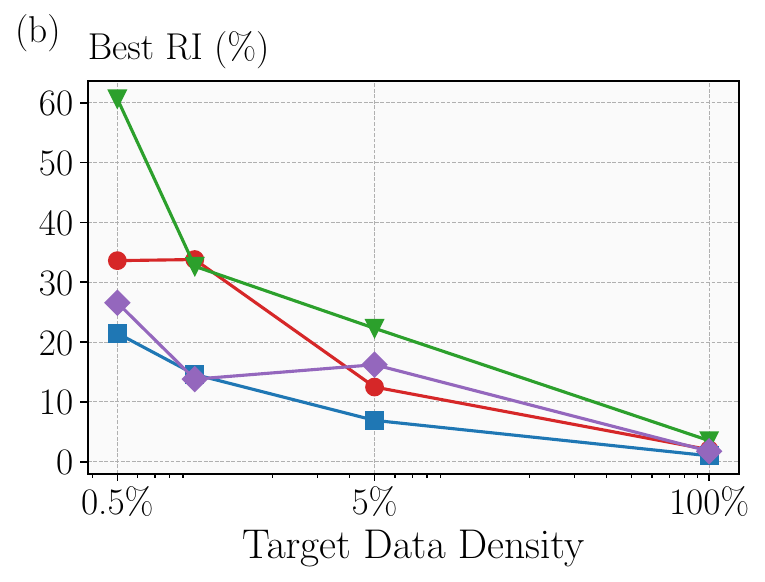}
\caption{Best RI from four source models versus target data density.}
\label{best_RI}
\vspace{-1em}
\end{figure}

\begin{figure}[tp]  
\centering
\includegraphics[width=0.84\linewidth]{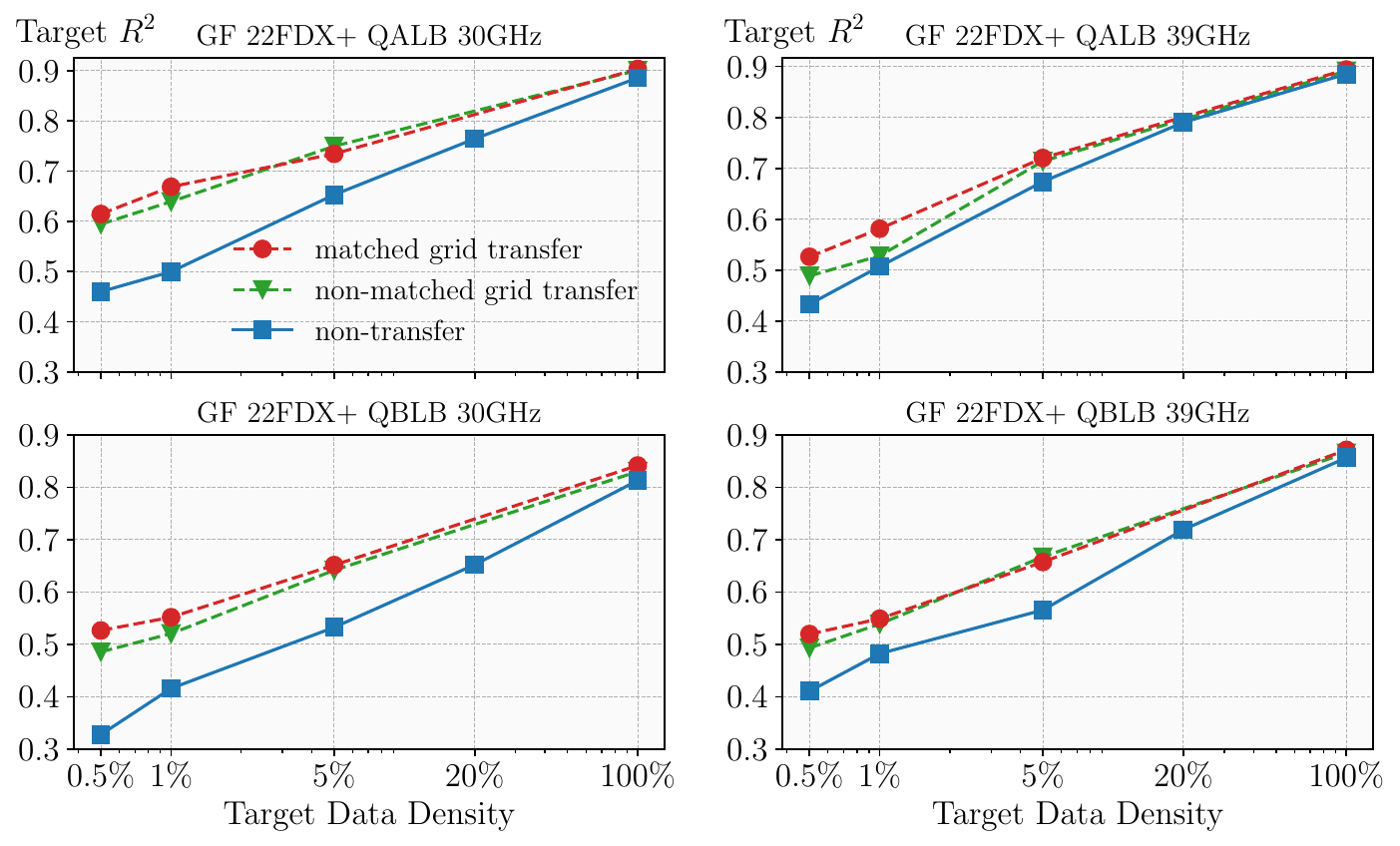}
\caption{Comparison of the target $R^2$ between the best transfer learning and without transfer.}
\label{best_Rsquare}
\vspace{-1.2em}
\end{figure}

Since the target grid has sparse data, training on the target grid is faster compared to the training source model itself. Therefore, we maintain multiple source models and transfer their knowledge to the target grid, respectively. Fig. \ref{best_RI}(a) shows the best RI from transferring the four source models in all settings, with the corresponding $R^2$ values displayed in \mbox{Fig. \ref{best_Rsquare}}(green dashed and blue solid). In general, the effect of transfer learning increases when the target data is more sparse. 

One remarkable observation is that the target data size can be reduced by a factor of 4, in GF 22FDX+ at 30GHz, for both QALB and QBLB metal options. This is evident as models trained with just 5\% of target data, enhanced by transfer learning, achieve higher $R^2$ values than those trained with 20\% without transfer, echoing the advantage seen from 1\% to 5\% data density. Likewise, for GF 22FDX+ QBLB 39GHz, $R^2$ values in models with sparse data and transfer learning are comparable to those in dense data models without transfer. The physical design parameters generated by the target models, augmented by transfer learning, will serve as a starting point for 3D EM fine-tuning of the on-chip transformer.

In practice, while source and target grids typically vary, there are scenarios where designers use the same grid for both. In these cases, data for GF 45SOI and GF 22FDX+ are generated on the same grid, leading to what we term the \emph{matched grid} scenario. Here, the source and target grids align, yet differ in technology nodes and operating frequencies. To accommodate this, we gather data in the source domain using an identical grid as the target domain. Fig. \ref{src}(b), Fig. \ref{best_RI}(b) and Fig. \ref{best_Rsquare} (red) display the source $R^2$, the best RI, and target $R^2$ for the matched grid case, respectively. As shown in \mbox{Fig. \ref{best_RI}}, the matched grid demonstrates superior transfer learning quality to the non-matched grid, due to a smaller distribution shift between the source and target domains. However, as revealed in Fig. \ref{best_RI}, these differences are modest, with the non-matched grid having a sufficiently high level of successful transfer learning. Thus, while the matched grid is optimal for better transfer learning, the non-matched grid performs sufficiently well as the source model for transfer learning.


\section{Conclusion}

We propose an innovative approach that utilizes transfer learning from a pre-trained synthesis NN model in one technology node and achieves swift and reliable design adaptation across different IC technologies, operating frequencies, and metal options. Simulation-based demonstrations prove that the transfer learning with the source domain knowledge (GF 45SOI) can not only accelerate the training process in the target domain (GF 22FDX+) but also improve the $R^2$ values compared to models without transfer learning. A notable reduction of 4X in the necessary target dataset size highlights the efficiency of transfer learning in mm-Wave passive network design. Consequently, this transfer learning approach promises to accelerate design migrations across technology nodes, thereby facilitating rapid time-to-market prototyping and productization.


\section*{Acknowledgment}
The authors thank GlobalFoundries for providing the information on different technology nodes. They also thank the members of the Integrated Devices, Electronics, And Systems (IDEAS) Group, ETH Z\"urich, for the technical discussions.


\bibliographystyle{IEEEtran}

\bibliography{IMS_Paper_LaTeX_Template_Letter_V6/IEEEexample}

\end{document}